\documentclass[12pt]{iopart}
\usepackage[slovene,english]{babel} 
\usepackage{graphicx,bbm,amssymb,amsopn}
\usepackage{dsfont}

\begin{document}

\title[Explicit solution of nearly isotropic boundary driven open XY spin 1/2 chain]{Explicit solution of the Lindblad equation for nearly isotropic boundary driven XY spin 1/2 chain}

\author{Bojan \v{Z}unkovi\v{c}${}^{1}$ and Toma\v{z} Prosen${}^{1,2}$ }

\address{${}^1$Department of Physics, Faculty of Mathematics and Physics, University of Ljubljana, Ljubljana, Slovenia\\
${}^2$ Department of Physics and Astronomy, University of Potsdam, Potsdam, Germany}

\date{\today}

\begin{abstract}
Explicit solution for the 2-point correlation function in a non-equilibrium steady state of a nearly isotropic boundary-driven open XY spin $1/2$ chain in the Lindblad
formulation is provided. A non-equilibrium quantum phase transition from exponentially decaying correlations to long-range order is discussed analytically. In the regime of long-range order a new phenomenon of {\em correlation resonances} is reported, where the correlation response of the system is unusually high for certain discrete values of the external bulk parameter, e.g. the magnetic field.
\end{abstract}

~~~~~~~~~~~~PACS numbers: 03.65.Yz, 02.30.Ik, 05.30.Fk, 75.10.Pq

\maketitle

\section{Introduction}
In recent years we are witnessing an increasing activity in non-equilibrium physics of interacting many-body problems. The reason may be at least two fold. Namely, on one hand this type of problems has nowadays become amenable to a detailed real lab experiments, e.g. in the context of cold atoms and optical lattices \cite{bloch}, and on the other hand, they are arguably connected to important open challenges in solid state physics, such as solid-state quantum computation or high temperature super-conductivity.
Essentially, there are two basic approaches to mathematical treatment of non-equilibrium many-body-physics. Perhaps more common approach \cite{general} is to consider a large (ideally infinite) system in an equilibrium state, then at some instant of time perform a quench of the Hamiltonian or join (couple) several infinite pieces of the system in distinct equilibria, and observe what happens in the course of time. For example, there might exist stationary long time behaviour. The other, more explicit approach \cite{alicki,breuer}, which we consider here, is to couple a finite (though perhaps large enough) system to several external reservoirs which we may describe {\em effectively} in terms of a master equation (by tracing out their explicit degrees of freedom), and consider a {\em non-equilibrium steady state} (NESS) to which a central system converges after a long time.

There is a variety of techniques involving several levels of assumptions and approximations in deriving the effective system's dynamics after tracing out the reservoirs degrees of freedom \cite{alicki,breuer}, which at the end result in a simple local-in-time (Markovian) linear differential equation for the system's density matrix, the so-called quantum Liouville equation. The more general Markovian form of such equations is sometimes referred to as the Redfield equation, whereas the more mathematically appealing form which manifestly conserves the positivity of the density matrix (and can be derived from the Redfield model with an additional, the so called {\em secular}  or {\em rotating wave} approximaiton, is the famous Lindblad equation \cite{lindblad}. It has been noted that Lindblad driven quantum chains provide a fruitful ground for studying non-equilibrium phase transitions \cite{zoller}.

Despite the fact that many simple cases of the Redfield and Lindblad master equations have been studied, a very few exactly solvable instances where the central system consists of many interacting particles have been discovered so far. In particular, in the context of addressing the quantum transport problem \cite{gemmer,wichterich}, one is interested in the quantum chain which is coupled to effective (thermal, chemical, or magnetic) reservoirs only at the ends of the chain (i.e. via the first and the last spin/particle).   

As an important simple but notable class of exactly solvable quantum Liouville equations one can identify open XY spin chains, or open quasi-free fermionic systems (to which XY chains can be mapped via Jordan-Wigner transformation) in general.
As it has been shown in \cite{njp}, the quantum Liouvillean of a general quadratic system of $n$ fermions can be explicitly diagonalized using the generalized (non-unitary) Bogoliubov-like transformation, and all the properties of NESS and the relaxation process can be computed explicitly in terms of diagonalization of a $4n\times 4n$ matrix.
The main conceptual tool in this approach is the introduction of the Fock-space structure over the space of operators which the density matrix is a member of.
This method has been later generalized also to more general Redfield master equations \cite{njp2} and put to a more rigorous mathematical footing \cite{pro10}.
It is remarkable that explicit solution for the NESS density matrix can be found also in some cases which go beyond quasi-free models, e.g. in XX spin 1/2 chain with on-site dephasing noise \cite{znidar} or in the XXZ spin 1/2 chain for the extremal (strong) driving \cite{prosaito}. On the side of numerical methods, the Liouville space version of the time-dependent density matrix renormalization group \cite{zwolak,daley} was successfully implemented to simulate steady states of open quantum chains \cite{proznid09}. 
For some other related explicit results on open quantum chains see e.g. Refs.\cite{other}.

As it has been shown in \cite{prl}, an open XY chain (either in the local Lindblad or Redfield \cite{njp2} setting) 
exhibits a quantum phase transition from exponentially decaying correlations to long range order in NESS when the
(bulk) parameters of the model are varied. Although a simple heuristic picture in terms of dispersion relation of the quasi-particle normal modes has been proposed to explain the
transition, the precise mathematical and physical understanding remained lacking.
In this paper we show that the non-equilibrium transition can be explained with explicit analytical calculation in the special regime of {\em small anisotropy}, where the perturbation theory in the anisotropy parameter can be successfully applied. We note that the small anisotropy open XY chain has been recently also studied numerically in terms of the Keldysh formalism \cite{marcin}, where the non-equilibrium phase transition has been re-confirmed.

Even though being surprisingly simple, our results analytically reproduce all the features of the transition to long range order under the additional assumption of weak coupling to the baths: namely we (i) give explicit expression for the 2-point correlation function, (ii) evaluate the decay rate of the correlation function in the short-range regime, and (iii) predict a new phenomenon in the long-range regime, the so called {\em resonances} of the correlation function at particular discrete values of the bulk parameter, e.g. the transverse magnetic field. At these resonant values of the magnetic field, in the long range order regime, the response of the system to external driving is particularly large as characterized by spontaneous emergence of 
strong correlations over large distances.

The paper is organized as follows. In section 2 we review the main concepts of quantization in the Fock space of operators, write down explicit equations of motion, and derive the so-called Lyapunov equation, as the dynamical equation for the 2-point correlation function which completely determines NESS. We stress that the results of this section represent certain simplification and improvements over the formulation of the previous papers \cite{njp,njp2}, namely we write for the first time a uniform quadratic form for the Liouvillean which remains valid in both, {\em even} and {\em odd parity} sectors of the operator space. This we do with a small trick, a simple redefinition of the canonical Majorana maps over the operator space.
In section 3 we then solve perturbatively, uniformly in two small parameters, the anisotropy and the system-bath coupling strength, the Lyapunov equation for the correlation (or covariance) matrix for the nearly isotropic open XY chain. We finally discuss our results and conclude in section 4.
 
\section{Equation of motion in operator Fock space}

In this section we make a very brief review of the technique of Liouvillean diagonalization of quadraric fermi systems in the Fock space of operators  \cite{njp,njp2} with an emphasis on the equation of motion and real-time dynamics. We treat a finite system with $n$ fermionic degrees of freedom, described by $2n$ anti-commuting hermitian operators $w_j$ ($j=1,2,\ldots 2n$).  The Redfield master equation (see e.g. \cite{breuer}) \footnote{We use units in which Planck's constant $\hbar=1.$}
\begin{equation}
\frac{\mathrm{d}\rho}{\mathrm{d}t}=\hat{\mathcal{L}}\rho:=-({\rm i~ad}H)\rho+\hat{\mathcal{D}}\rho
\end{equation} 
serves us as the most general (Markovian) equation of motion that we are able to treat \cite{njp2} (quantum Liouville equation). The first term is anti-hermitian
$-({\rm i~ad}H)\rho:=-{\rm i}[H,\rho]$ and generates the usual unitary von Neumann equation. The second term - the {\em dissipator map} - has a memoryless kernel with the following general form \cite{breuer}
\begin{equation}
\label{redfieldeq}
\hat{\mathcal{D}}=\sum_{\mu,\nu}\int_0^\infty\!\!{\rm d}t\,\Gamma^\beta_{\nu,\mu}(t)[\tilde{X}_{\mu}(-t)\rho,X_\nu]+\mathrm{h.c.},
\end{equation}
where $\Gamma^\beta_{\nu,\mu}(t)$  denotes the environment (bath) correlation function and $\tilde{X}_\mu(t)$ denotes the Heisenberg picture of hermitian coupling operators $X_\mu$ (following notation of \cite{njp2}).

The Hamiltonian can be a general quadratic form in $w_j$ and the coupling operators should be linear in $w_j$
\begin{eqnarray}
H=\sum_{j,k=1}^nw_j\mathbf{H}_{j,k}w_k=\underline{w}\cdot\mathbf{H}\underline{w},\\ \nonumber
X_\mu =\sum_j x_{\mu,j} w_j=\underline{x}_\mu\cdot\underline{w}.
\end{eqnarray}
The $2n\times2n$ matrix $\mathbf{H}$ is antisymmetric. Throughout this paper $\underline{x} = (x_1, x_2,\ldots)^{\rm T}$ will designate a vector (column) of appropriate scalar valued or operator valued symbols $x_k$. 

\subsection{Bilinear form of the complete Liouvillean}
Diagonalization of the Liouvillean $\hat{\mathcal{L}}$ is done in the $4^n$ dimensional Liouville space of operators $\mathcal{K}$, which has the structure of the Fock space with the canonical basis
\begin{equation}
P_{\underline{\alpha}}=2^{-n/2}w_1^{\alpha_1}w_2^{\alpha_2}\ldots w_{2n}^{\alpha_{2n}},\quad \alpha_j \in \{ 0,1\}.
\end{equation}
We will use the bra-ket notation for the operator space $\mathcal{K}$ with the inner product 
\begin{equation}
\langle x|y\rangle=\mathrm{tr}(x^\dag y).
\end{equation} 
Further we define the creation/annihilation maps by $\hat{c}_i^\dag|P_{\underline{\alpha}}\rangle=(1-\alpha_j)|w_jP_{\underline{\alpha}}\rangle$,  $\hat{c}_i|P_{\underline{\alpha}}\rangle=\alpha_j|w_jP_{\underline{\alpha}}\rangle$, which satisfy {\em canonical anti-commutation relations} (CAR), $\{\hat{c}_i,\hat{c}^\dag_j\}=\delta_{i,j}$, $\{\hat{c}_i^\dag,\hat{c}^\dag_j\}=\{\hat{c}_i,\hat{c}_j\}=0$. The Redfield dissipator can be written as a quadratic form in these maps
\begin{equation}
\label{rdis:eq}
\hat{\mathcal{D}}=\sum_\nu\sum_{k,j=1}^{2n}x_{\nu,k}\left(z_{\nu,j}\hat{\mathcal{L}}'_{j,k}+ z_{\nu,j}^* \hat{\mathcal{L}}''_{j,k}\right),
\end{equation}
where $z_{\nu,j}$ are complex constants defined as
\begin{equation}
\underline{z}_\nu:=\sum_\mu\int_0^\infty\mathrm{d}t \exp(-4\mathrm{i}\mathbf{H}t)\underline{x}_\mu\Gamma^\beta_{\nu,\mu}(t)
\end{equation}
and
\begin{equation}
\hat{\mathcal{L}}'_{j,k}|x\rangle:=|[w_jx,w_k]\rangle,\quad \hat{\mathcal{L}}''_{j,k}|x\rangle:=|[w_k,xw_j]\rangle
\end{equation}
are fundamental basis dissipators which evaluate to
\begin{eqnarray}
\label{baisis_dis:eq}
\hat{\mathcal{L}}'_{j,k} &= \left( \hat{\mathds{I}}+\hat{\mathcal{P}}\right)(\hat{c}_j^\dag\hat{c}_k^\dag-\hat{c}_k^\dag\hat{c}_j)+\left( \hat{\mathds{I}}-\hat{\mathcal{P}}\right)(\hat{c}_j\hat{c}_k-\hat{c}_k\hat{c}_j^\dag), \\ \nonumber
\hat{\mathcal{L}}''_{j,k} &=\left( \hat{\mathds{I}}+\hat{\mathcal{P}}\right)(\hat{c}_k^\dag\hat{c}_j^\dag-\hat{c}_k^\dag\hat{c}_j)+\left( \hat{\mathds{I}}-\hat{\mathcal{P}}\right)(\hat{c}_k\hat{c}_j-\hat{c}_k\hat{c}_j^\dag).
\end{eqnarray}
Note that the parity map $\hat{\mathcal{P}}=\exp\left(\mathrm{i}\pi\sum_{j=1}^{2n}\hat{c}_j^\dag\hat{c}_j \right)$ and the positive and negative parity projectors $\hat{\mathcal{P}}^\pm=\frac{1}{2}(\hat{\mathds{I}}\pm\hat{\mathcal{P}})$ commute with the Liouvillean, $[\hat{\mathcal{L}},\hat{\mathcal{P}}^{(\pm)}]=0$. Plugging the definitions (\ref{baisis_dis:eq}) in the equation for the Redfield map (\ref{rdis:eq}) we see that the dissipator decomposes into a positive ($\hat{\mathcal{D}}^+$) and a negative ($\hat{\mathcal{D}}^-$) parity components
\begin{eqnarray}
\label{disipL:eq}
\hat{\mathcal{D}}&=&\hat{\mathcal{D}}^+\hat{\mathcal{P}}^++\hat{\mathcal{D}}^-\hat{\mathcal{P}}^-,\\ \nonumber
\hat{\mathcal{D}}^+&=&2\underline{\hat{c}}^\dag\cdot\left({\bf M}^{\rm T}+{\bf M}^*\right)\underline{\hat{c}}^\dag-2\underline{\hat{c}}^\dag\cdot\left({\bf M}+{\bf M}^*\right)\underline{\hat{c}},\\ \nonumber
\hat{\mathcal{D}}^-&=&2\underline{\hat{c}}\cdot\left({\bf M}^{\rm T}+{\bf M}^*\right)\underline{\hat{c}}-2\underline{\hat{c}}\cdot\left({\bf M}+{\bf M}^*\right)\underline{\hat{c}}^\dag,
\end{eqnarray}
where ${\bf M}$ is a {\em bath-matrix} which can be compactly written as ${\bf M}:=\sum_\nu \underline{x}_\nu\otimes \underline{z}_\nu$, and ${\bf M}^*$ and ${\bf M}^{\rm T}$ denote respectively, the complex conjugate and transpose of the matrix. The Liouvillean can be written in a convenient form using $4n$ fermionic Majorana maps
\begin{eqnarray}
\hat{a}^{\circ}_{1,j}:=\frac{1}{\sqrt{2}}(\hat{c}_j+\hat{c}^\dag_j),\quad\hat{a}^{\circ}_{2,j}:=\frac{\mathrm{i}}{\sqrt{2}}(\hat{c}_j-\hat{c}^\dag_j)\quad\mbox{ and }\quad\hat{\underline{a}}^{\circ}=(\hat{\underline{a}}_{1}^{\circ},\hat{\underline{a}}_{2}^{\circ})
\end{eqnarray}
satisfying CAR $\{\hat{a}^{\circ}_{\mu,j},\hat{a}^{\circ}_{\nu,k}\}=\delta_{\mu,\nu}\delta_{j,k}$.
Straightforward calculation shows that the dissipator has the following form
\begin{eqnarray}
\hat{\mathcal{D}}^+=\hat{\underline{a}}_1^{\circ}\cdot{\bf A}^{\rm D}_{1,1}\hat{\underline{a}}_1^{\circ}+\hat{\underline{a}}_2^0\cdot{\bf A}^{\rm D}_{2,2}\hat{\underline{a}}_2^{\circ}+\hat{\underline{a}}_1^{\circ}\cdot{\bf A}^{\rm D}_{1,2}\hat{\underline{a}}_2^{\circ}+\hat{\underline{a}}_2^{\circ}\cdot{\bf A}^{\rm D}_{2,1}\hat{\underline{a}}_1^{\circ}-A_0\hat{\mathds{I}},\\ \nonumber
\hat{\mathcal{D}}^-=\hat{\underline{a}}_1^{\circ}\cdot{\bf A}^{\rm D}_{1,1}\hat{\underline{a}}_1^{\circ}+\hat{\underline{a}}_2^{\circ}\cdot{\bf A}^{\rm D}_{2,2}\hat{\underline{a}}_2^{\circ}-\hat{\underline{a}}_1^{\circ}\cdot{\bf A}^{\rm D}_{1,2}\hat{\underline{a}}_2^{\circ}-\hat{\underline{a}}_2^{\circ}\cdot{\bf A}^{\rm D}_{2,1}\hat{\underline{a}}_1^{\circ}-A_0\hat{\mathds{I}},
\end{eqnarray}
where we used the definitions
\begin{eqnarray}
&{\bf A}_{1,1}^{\rm D}:={\bf M}^{\rm T}-{\bf M},&\quad{\bf A}_{1,2}^{\rm D}:={\rm i}{\bf M}^{\rm T}+{\rm i}{\bf M}^*,\\ \nonumber
& {\bf A}_{2,1}^{\rm D}:=-{\rm i}{\bf M}^{\dag}-{\rm i}{\bf M},&\quad {\bf A}_{2,2}^{\rm D}:={\bf M}^{\dag}-{\bf M}^*,\\ \nonumber
&A_0 := {\rm tr}({\bf M+M^*}).&
\end{eqnarray}
The unitary part of the Liouvillean is also quadratic in the  Majorana maps $\hat{\underline{a}}^{\circ}_{1},\hat{\underline{a}}^{\circ}_{2}$ \cite{njp,njp2}
\begin{eqnarray}
\label{unitL:eq}
-{\rm i~ ad} H=-4{\rm i}\hat{\underline{c}}^\dag\cdot{\bf H}\hat{\underline{c}}=-2{\rm i}(\hat{\underline{a}}_1^{\circ}\cdot{\bf H}\hat{\underline{a}}_1^{\circ}+\hat{\underline{a}}_2^{\circ}\cdot{\bf H}\hat{\underline{a}}_2^{\circ}).
\end{eqnarray}
Collecting the results (\ref{disipL:eq}), (\ref{unitL:eq}) the complete Liouvillean decomposes in the same way as its dissipative part
\begin{eqnarray}
\label{liouvproj:eq}
\hat{\mathcal{L}}&=&\hat{\mathcal{L}}^+\hat{\mathcal{P}}^++\hat{\mathcal{L}}^-\hat{\mathcal{P}}^-,\\ \nonumber
\hat{\mathcal{L}}^+&=&\hat{\underline{a}}_1^{\circ}\cdot{\bf A}_{1,1}\hat{\underline{a}}_1^{\circ}+\hat{\underline{a}}_2^{\circ}\cdot{\bf A}_{2,2}\hat{\underline{a}}_2^{\circ}+\hat{\underline{a}}_1^{\circ}\cdot{\bf A}_{1,2}\hat{\underline{a}}_2^{\circ}+\hat{\underline{a}}_2^{\circ}\cdot{\bf A}_{2,1}\hat{\underline{a}}_1^{\circ}-A_0\hat{\mathds{I}},\\ \nonumber
\hat{\mathcal{L}}^-&=&\hat{\underline{a}}_1^{\circ}\cdot{\bf A}_{1,1}\hat{\underline{a}}_1^{\circ}+\hat{\underline{a}}_2^{\circ}\cdot{\bf A}_{2,2}\hat{\underline{a}}_2^{\circ}-\hat{\underline{a}}_1^{\circ}\cdot{\bf A}_{1,2}\hat{\underline{a}}_2^{\circ}-\hat{\underline{a}}_2^{\circ}\cdot{\bf A}_{2,1}\hat{\underline{a}}_1^{\circ}-A_0\hat{\mathds{I}},
\end{eqnarray}
where we introduce $2n\times 2n$ matrices
\begin{equation}
{\bf A}_{\mu,\nu} := -2{\rm i}\delta_{\mu,\nu}{\bf H} + {\bf A}^{\rm D}_{\mu,\nu},\quad \mu,\nu=1,2.
\end{equation}
Summing the odd and even parity Liouvilleans (\ref{liouvproj:eq}) and using the definition of the parity projectors we obtain
\begin{equation}
\label{sfliouv:eq}
\hat{\mathcal{L}}=\hat{\underline{a}}_1^{\circ}\cdot{\bf A}_{1,1}\hat{\underline{a}}_1^{\circ}+\hat{\underline{a}}_2^{\circ}\cdot{\bf A}_{2,2}\hat{\underline{a}}_2^{\circ}+(\hat{\underline{a}}_1^{\circ}\cdot{\bf A}_{1,2}\hat{\underline{a}}_2^{\circ}+\hat{\underline{a}}_2^{\circ}\cdot{\bf A}_{2,1}\hat{\underline{a}}_1^{\circ})\hat{\mathcal{P}}-A_0\hat{\mathds{I}}.
\end{equation}
To get rid of the parity map in the Liouvillean (\ref{sfliouv:eq}) we define modified hermitian Majorana maps  
\begin{equation}
\label{amapfine:eq}
\hat{a}_{1,j}:={\rm i}\hat{\eta} \hat{a}_{1,j}^{\circ},\quad \hat{a}_{2,j}:={\rm i}\hat{\eta} \hat{a}_{2,j}^{\circ}\hat{\mathcal{P}},\quad  \hat{\underline{a}}=(\hat{\underline{a}}_1,\hat{\underline{a}}_2),
\end{equation}
where $\hat{\eta}=2^n{\rm i}\prod_{j=1}^{2n}\hat{a}^\circ_{1,j}$ is a hermitian map obeying the following (anti)commutation relations
\begin{eqnarray}
\{\hat{a}_{1,j}^\circ,\hat{\eta}\}=0,\quad,[\hat{a}_{2,j}^\circ,\hat{\eta}]=0\quad [\hat{\eta},\hat{\mathcal{P}}]=0.
\end{eqnarray}
We note that new Majorana maps again satisfy CAR, $\{\hat{a}_{\mu,j},\hat{a}_{\nu,k}\} = \delta_{\mu,\nu}\delta_{j,k}$.
The Liouvillean takes now a simple quadratic form    
\begin{equation}
\label{liofine:eq}
\hat{\mathcal{L}}=\hat{\underline{a}}\cdot{\bf A}\hat{\underline{a}}-A_0\hat{\mathds{I}},
\end{equation}
where a structure matrix ${\bf A}$ is a block matrix
\begin{equation}
{\bf A}=\left(\begin{array}{cc}
{\bf A}_{1,1},&{\bf A}_{1,2}\\
{\bf A}_{2,1},&{\bf A}_{2,2}
\end{array}\right).
\end{equation}

Incorporating the odd subspace to the Liouvillean enables us to treat the time evolution of a general density operator with the formalism of  `third quantization'. 
Note that due to non-normality of the Liouvillean as an operator we have to distinguish between left and right vacua
\begin{equation}
\langle 1|\hat{\mathcal{L}}=0\quad\mbox{and}\quad \hat{\mathcal{L}}|{\rm NESS}\rangle=0.
\end{equation} 
The left vacuum is simply the identity operator, whereas the right vacuum represents a density operator of NESS.

\subsection{Derivation of the Lyapunov equation}
Taking any time dependent solution of quantum Liouville equation $\rho(t)$ it is convenient to study its 2-point correlation function encoded in the $2n\times 2n$ correlation or covariance matrix
\begin{equation}
C_{j,k}(t)=\mathrm{tr}\left(w_j w_k\rho(t)\right)-\delta_{j,k},
\end{equation}
which may be written compactly in terms of the Majorana maps (\ref{amapfine:eq})
as
\begin{equation}
{\bf C}(t)=2\langle 1|\hat{\underline{a}}_1\otimes\hat{\underline{a}}_{1}|\rho(t)\rangle-\mathds{I}_{2n}.
\end {equation}
Note that in the steady state, since NESS is a generalized Gaussian sate, all the information about NESS can be extracted from ${\bf C}$ by means of the Wick theorem.
Expressing the stationarity of the left vacuum $(\langle 1|\exp(\hat{\mathcal{L}}t)=\langle 1|)$ and using the trivial identities\footnote{We assume that the Hamiltonian and the coupling operators $\hat{X}_\mu$ are time independent.}
\begin{equation}
{\rm e}^{\hat{\mathcal{L}}t}{\rm e}^{-\hat{\mathcal{L}}t}=\hat{\mathds{I}},\quad\quad
|\rho(t)\rangle={\rm e}^{\hat{\mathcal{L}}t}|\rho(0)\rangle
\end{equation}
the correlation matrix can be further simplified 
\begin{eqnarray}
\label{corrmatrixeq}
C_{j,k}(t)+\delta_{j,k}&=2\langle 1|\hat{a}_{1,j}\hat{a}_{1,k}{\rm e}^{\mathcal{L}t}|\rho(0)\rangle=2\langle 1|{\rm e}^{-\hat{\mathcal{L}}t}\hat{a}_{1,j}{\rm e}^{\hat{\mathcal{L}}t}{\rm e}^{-\hat{\mathcal{L}}t}\hat{a}_{1,k}{\rm e}^{\hat{\mathcal{L}}t}|\rho(0)\rangle\\ \nonumber
&=2\langle 1|\hat{a}_{1,j}(t)\hat{a}_k(t)|\rho(0)\rangle,
\end{eqnarray}
where a super-Heisenberg picture is defined $\hat{a}_i(t):={\rm e}^{-\hat{\mathcal{L}}t}\hat{a}_i{\rm e}^{\hat{\mathcal{L}}t}$. Using quadratic form of the Liouvillean (\ref{liofine:eq}), we get the Heisenberg equation of motion for the Majorana maps
\begin{equation}
\frac{{\rm d}\hat{a}_{\nu,j}(t)}{{\rm d}t}=[\hat{a}_{\nu,j}(t),\hat{\mathcal{L}}]=2\sum_{\mu=1}^2 \sum_{l=1}^{2n}A_{(\nu,j),(\mu,l)}\hat{a}_{\mu,l}(t),
\end{equation}
where a multi-index $(\nu,j)$ is used with $\nu=1,2$ and $j=1,2,\ldots2n$. Differentiating (\ref{corrmatrixeq}) with respect to time gives
\begin{eqnarray}
\label{strmatrix:eq}
\frac{{\rm d} C_{j,k}}{{\rm d}t}=4\sum_{\mu=\{1,2\}}\sum_{l=1}^{2n}&\Big(&A_{(1,j),(\mu,l)}\langle 1|\hat{a}_{\mu,l}(t)\hat{a}_{1,k}(t)|\rho(0)\rangle
\\ \nonumber &+& A_{(1,k),(\mu,l)}\langle 1|\hat{a}_{1,j}(t)\hat{a}_{\mu,l}(t)|\rho(0)\rangle\Big).
\end{eqnarray}
Taking into account the identities (assuming $l\neq k$)
\begin{eqnarray}
&2\langle 1|\hat{a}_{1,l}(t)\hat{a}_{1,k}(t)|\rho(0)\rangle= C_{l,k}(t),&\quad 2\langle 1|\hat{a}_{2,l}(t)\hat{a}_{2,k}(t)|\rho(0)\rangle=- C_{l,k}(t),\\ \nonumber
&2\langle 1|\hat{a}_{1,l}(t)\hat{a}_{2,k}(t)|\rho(0)\rangle={\rm i} C_{l,k}(t),&\quad 2\langle 1|\hat{a}_{2,l}(t)\hat{a}_{1,k}(t)|\rho(0)\rangle={\rm i}C_{l,k}(t),\\ \nonumber
&2\langle 1|\hat{a}_{2,k}(t)\hat{a}_{1,k}(t)|\rho(0)\rangle=-{\rm i},&\quad 2\langle 1|\hat{a}_{1,k}(t)\hat{a}_{2,k}(t)|\rho(0)\rangle={\rm i},
\end{eqnarray}
and the definition of ${\bf A}$ we derive the differential equation for the time-dependent correlation matrix
\begin{eqnarray}
\label{correq}
\frac{\mathrm{d}{\bf C}}{\mathrm{d}t} = 
-4\mathrm{i} [\mathbf{H,C}]-4(\mathbf{M}_\mathrm{r} \mathbf{C}+\mathbf{C} \mathbf{M}_\mathrm{r}^{\rm T})-4{\rm i}(\mathbf{M}_{\rm i}-\mathbf{M}^{\rm T}_{\rm i}),
\end{eqnarray}
where $\mathbf{M}_\mathrm{r}:=\mathrm{Re}(\mathbf{M})=(\mathbf{M}+\mathbf{M}^*)/2$, $\mathbf{M}_{\rm i} := \mathrm{Im}(\mathbf{M}) = (\mathbf{M}-\mathbf{M}^*)/(2{\rm i})$. Note again that the stationary solution $\frac{\mathrm{d}\mathbf{C}}{\mathrm{d}t}=0$ fully determines NESS. The equation for the  stationary correlation matrix, obtained from  (\ref{correq}), can be written compactly using the anti-symmetry of Hamiltonian matrix $\mathbf{H}$ 
\begin{eqnarray}
\mathbf{C}\mathbf{X}^{\rm T}+\mathbf{XC}=\mathbf{Y},
\label{correq1}
\end{eqnarray}  
where new matrices $\mathbf{X}:=4({\rm i}\mathbf{H}+\mathbf{M}_\mathrm{r})$ and $\mathbf{Y}:=4{\rm i}(\mathbf{M}_{\rm i}^{\rm T}-\mathbf{M}_{\rm i})$ are defined. This 
matrix equation is known as the continuous Lyapunov equation in control theory and stability analysis and has a unique solution if and only if the eigenvalues 
$\beta_j$ of the matrix $\mathbf{X}$ satisfy \cite{lyapunov} 
\begin{equation}
\label{conditioneq}
\beta_j+\beta_k\neq0,\quad j,k=1,2,\ldots n.
\end{equation}

We note that although using a general Redfield form of the dissipator the results are valid also for the case of Lindblad equation (\ref{lindbladeq}), as the most general Lindblad dissipator can be obtained from (\ref{redfieldeq}) by taking the delta-like bath correlation function $\Gamma^\beta_{\mu,\nu}(t)=\gamma_{\mu,\nu}\delta(t+0)$ \cite{breuer,njp2}. 
In the next section we shall apply these theoretical considerations precisely in the case of Lindblad equation.

The Lyapunov equation (\ref{correq1}) has theoretical and practical implications. On the theoretical side, it helps to understand the conditions under which non-equilibrium stationary state is unique, it is related to the full spectrum of the structure matrix $\mathbf{A}$, and simplifies the analytical calculations, while for practical purposes significantly simplifies the numerical implementation of the method of `third quantization', reduces the matrix dimension needed in the computations, and improves the stability of the method.

\section{Nearly isotropic XY model}

In the remainder of the paper we consider the Lindblad equation
\begin{eqnarray}
\label{lindbladeq}
\frac{\mathrm{d}\rho}{{\rm d}t}=-\mathrm{i}[H,\rho]+\sum_\mu(2L_\mu\rho L_\mu^\dag-\{L_\mu^\dag L_\mu,\rho\})
\end{eqnarray}
for an open XY spin 1/2 chain with the Hamiltonian
\begin{eqnarray}\nonumber
H&=\sum_{m=1}^{n-1}\Big(\frac{1-\gamma}{2}\sigma_{m}^{\rm x}\sigma_{m+1}^{\rm x}+\frac{1+\gamma}{2}\sigma_{m}^{\rm y}\sigma_{m+1}^{\rm y}\Big) +\sum^n_{m=1}h\sigma_{m}^{\rm z},
\end{eqnarray}
with four Lindblad operators coupling minimally to the ends of the chain
\begin{equation}
L_{1,2} = \sqrt{\Gamma^{\rm L}_{1,2}} \sigma^{\pm}_1,
\quad
L_{3,4} = \sqrt{\Gamma^{\rm R}_{1,2}} \sigma^{\pm}_n.
\end{equation}
Using the Jordan-Wigner transformation the XY model is mapped to a fermionic model with the Hamiltonian
\begin{eqnarray}\nonumber
H&=-\mathrm{i}\sum_{m=1}^{n-1}\Big(\frac{1-\gamma}{2}w_{2m}w_{2m+1}+\frac{1+\gamma}{2}w_{2m-1}w_{2m+2}\Big) - \mathrm{i}\sum^n_{m=1}hw_{2m-1}w_{2m}\\
&=:\underline{w}\cdot \mathbf{H} \underline{w}.
\end{eqnarray} 
The equivalent Lindblad operators\footnote{The non-local Jordan-Wigner transformation brings a non-local Casimir operator and a phase factor in front of $L_{3,4}$ which however do not alter the Lindblad equation (\ref{lindbladeq}).} are linear in Majorana fermions
\begin{eqnarray}
L_{1,2}&=\frac{1}{2}\sqrt{\Gamma^{\rm L}_{1,2}}(w_1\pm {\rm i}w_2)=\underline{l}_{1,2}\cdot\underline{w},\\ \nonumber
L_{3,4}&=\frac{1}{2}\sqrt{\Gamma^{\rm R}_{1,2}}(w_{2n-1}\pm {\rm i}w_{2n})=\underline{l}_{3,4}\cdot\underline{w}.
\end{eqnarray}
It is convenient to define new coupling constants as $\Gamma^{\rm L,R}_{1,2}=:\epsilon \kappa_{1,2}^{\rm L,R}$,
introducing a {\em small parameter} $\epsilon$. We also define 
\begin{equation}
	\Gamma^{\rm L,R}_{\pm}=\Gamma^{\rm L,R}_{1}\pm\Gamma^{\rm L,R}_{2},\quad 	\kappa^{\rm L,R}_{\pm}=\kappa^{\rm L,R}_{1}\pm\kappa^{\rm L,R}_{2}.
\end{equation}

\subsection{Asymptotic analytic solution}

The analytic expression for the correlation function of weakly coupled nearly isotropic XY spin 1/2 model can be derived in two steps. Firstly, we make an ansatz for the correlation matrix 
\begin{equation}
\mathbf{C}=\sum_{j,k}\Lambda_{j,k}\underline{\Psi}_j^\mathrm{right}\otimes\underline{\Psi}_k^\mathrm{right},
\end{equation} 
where $\underline{\Psi}_j^\mathrm{right}$ denotes the $j$-th right eigenvector of $\mathbf{X}$ with the corresponding eigenvalue $\beta_j$,
$\mathbf{X}\underline{\Psi}_j^\mathrm{right}=\beta_j\underline{\Psi}_j^\mathrm{right}$. Then, by plugging the ansatz into the equation (\ref{correq1}) we get the expression for the coefficients $\Lambda_{j,k}$
\begin{equation}
\Lambda_{j,k}=\frac{1}{\beta_j+\beta_k}\underline{\Psi}_j^{\mathrm{left}*}\cdot\mathbf{Y}\underline{\Psi}_k^{\mathrm{left}*}.
\label{coefeq}
\end{equation}
Left and right eigenvectors are normalized as $\underline{\Psi}_j^{\mathrm{left}*}\cdot\underline{\Psi}_k^\mathrm{right}=\delta_{j,k}.$ 

Secondly, we find approximate eigenvectors of $\mathbf{X}$ and determine the coefficients $\Lambda_{j,k}$ for nearly isotropic XY spin 1/2 chain with Lindbald reservoirs. Explicit solution is found by means of perturbation theory using two main assumptions: (i) small anisotropy ($\gamma\ll1$), and (ii) weak coupling ($\epsilon\ll\gamma$).  Hence, the matrix $\mathbf{X}$ can be decomposed into three parts
\begin{equation}
{\bf X}={\bf X}_0+\gamma{\bf X}_\gamma+\epsilon{\bf X}_\epsilon.
\end{equation}
The first one is essentially the (isotropic) XX Hamiltonian, the second one is its anisotropic part, and the last one comes from the coupling to the environment ($4\mathbf{M}_{\rm r}=\epsilon{\bf X}_\epsilon$)  
\begin{eqnarray}
{\bf X}_0=2{\rm i}h\sigma^{\rm y}\otimes\mathds{I}_{n}-{\rm i}\sigma^{\rm y}\otimes{\bf J}_{n}, \\ \nonumber
{\bf X}_\gamma=\sigma^{\rm x}\otimes{\bf J}'_{n}, \\ \nonumber
{\bf X}_\epsilon=
\left(
\begin{array}{cccc}
\kappa^{\rm L}_+\mathds{I}_{2} & 0 & \ldots & 0 \\
 0 & 0 & \ldots & 0 \\
 \vdots & \vdots & \ddots &  \vdots \\
 0  & 0 & \ldots & \kappa^{\rm R}_+\mathds{I}_{2}
\end{array}
\right),
\end{eqnarray}
where $\mathds{I}_n$ is a $n\times n$ identity matrix and ${\bf J}_n$ and ${\bf J}'_n$ are $n\times n$ matrices
\begin{eqnarray}
{\bf J}_{n}:=\left(
\begin{array}{ccccc}
 0 & 1 & 0 & 0 & \ldots\\
 1 & 0 & 1 & 0 & \ldots\\
 0 & 1 & 0 &  1 & \ldots\\
 0  & 0 & 1 &0 & \ldots\\
 \vdots & \vdots & \vdots & \vdots & \ddots
\end{array}
\right),\quad 
{\bf J}'_{n}:=\left(
\begin{array}{ccccc}
 0 & 1 & 0 & 0 & \ldots\\
 -1 & 0 & 1 & 0 & \ldots\\
 0 & -1 & 0 &  1 & \ldots\\
 0  & 0 & -1 &0 & \ldots\\
 \vdots & \vdots & \vdots & \vdots & \ddots
\end{array}
\right).
\end{eqnarray}
Note that the matrices $\mathbf{X}_0, \mathbf{X}_\gamma$ are anti-hermitian and real (real and antisymmetric), while the matrix $\mathbf{X}_\epsilon$ is hermitian (real and symmetric). Exact eigenvectors and eigenvalues for such system are rather difficult to calculate, therefore, we use the perturbation theory to incorporate the effects of the environment and the anisotropy. The unperturbed term (${\bf X}_0$) is exactly diagonalizable by the Fourier transformation, namely $\mathbf{X}_0 \underline{\Psi}^0_{\omega,j} = \beta^0_{\omega,j} \underline{\Psi}^0_{\omega,j}$ with
\begin{eqnarray}
&\underline{\Psi}_{\omega,j}^0=\frac{(1,\pm\mathrm{i})}{\sqrt{2}}\otimes\underline{\psi_j},&
\quad \beta^0_{\omega,j}=\pm 2\mathrm{i}  \left(h-\cos \left(\frac{\pi  j}{n+1}\right)\right),
\label{eq:beta0}
\end{eqnarray}
where a multi-index $(\omega,j)$, $\omega=1,2$, $j=1,2,\ldots,n$ is introduced in order to label eigenvalues/eigenvectors, and the plus ($+$), or minus ($-$) sign,
in (\ref{eq:beta0}) corresponds to $\omega=1$, or $\omega=2$, respectively, and  
 \begin{equation}
 \psi_{j,k}=\sqrt{\frac{2}{n+1}}\sin\left[\frac{\pi j k}{n+1}\right].
 \label{eq:sw}
 \end{equation}
 
We can neglect the effect of environment to the eigenvectors in the case of very small coupling $\epsilon\ll\gamma\ll1$ 
\begin{equation}
\underline{\Psi}_{\omega,j}=\underline{\Psi}^0_{\omega,j}+\gamma \underline{\Psi}^1_{\omega,j}+\mathcal{O}(\epsilon)+\mathcal{O}(\gamma^2),
\end{equation}
but we must include the first correction to the eigenvalues 
\begin{equation}
\beta_{\omega,j}=\beta^0_{\omega,j}+\epsilon \beta^1_{\omega,j}+\mathcal{O}(\epsilon^2)+\mathcal{O}(\gamma),
\end{equation}
where, on the contrary, the effects of the anisotropy can be neglected to leading order. Reasons for this asymmetry will become clear later on when we will calculate the coefficients $\Lambda_{j,k}$.  Thus, in case of the eigenvalues we need to calculate only the real (dissipative) corrections to first order in $\epsilon$, although the imaginary corrections from the anti-hermitian part can be much larger (as we have assumed $\epsilon\ll\gamma$)
\begin{eqnarray}
\beta_{\omega,j}^1=
\frac{2}{ (n+1)}
\left[\kappa^\mathrm{L}_++\kappa^\mathrm{R}_+ \right] \sin ^2\left(\frac{\pi  j}{n+1}\right), \quad\omega=1,2.
\end{eqnarray}
Note that the correction is positive for both, positive and negative imaginary unperturbed eigenvalues (energies) $\mathrm{Im}(\beta_{\omega,j}^0)$. In contrast to eigenvalues, as already noted, important corrections to eigenvectors come from the anti-hermitian part ${\bf X}_\gamma$ (because of the assumption $\epsilon\ll\gamma$) and are calculated by the usual first order perturbation theory
\begin{eqnarray}
\underline{\Psi}_{2,j}^1=\gamma\sum_{j'=1}^n\frac{K_{j,j'}}{\beta_{2,j}^0-\beta_{1,j'}^0}\underline{\Psi}_{1,j'}^0,\quad \underline{\Psi}_{1,j}^1=\gamma\sum_{j'=1}^n\frac{-K_{j,j'}}{\beta_{1,j}^0-\beta_{2,j'}^0}\underline{\Psi}_{2,j'}^0,
\end{eqnarray}
with
\begin{eqnarray}
K_{j,j'}=\underline{\Psi}_{2,j}^{0*}\cdot \mathbf{X}_\gamma \underline{\Psi}_{1,j'}^0
=\frac{\mathrm{i} \left(1-(-1)^{j+j'}\right)}{(n+1)}
\frac{\sin \left(\frac{\pi 
   j}{n+1}\right) \sin \left(\frac{\pi  j'}{n+1}\right)}{\sin
   \left(\frac{\pi  (j'-j)}{2 (n+1)}\right) \sin \left(\frac{\pi 
   (j+j')}{2 (n+1)}\right)}.
 \label{eq:K}
\end{eqnarray}
Long but straightforward calculation shows that all the matrix elements of ${\bf X}_\gamma$ connecting the vectors with the same polarization (i.e. vectors $\underline{\Psi}^0_{\omega,j}$ and $\underline{\Psi}^0_{\omega,j'}$ for $j',j=1,2\ldots n$,
 and $\omega=1,2$) vanish. Neglecting environment correction of eigenvectors has a convenient consequence, namely, the approximate left and right eigenvectors of the matrix $\mathbf{X}$ are identical. Note, importantly, that in this approach the anisotropy $\gamma$ must be smaller than $\delta\beta$, $|\gamma| < \delta\beta$, where $\delta\beta$ is the smallest spacing between adjacent unperturbed eigenvalues $\beta^0_{\omega,j}$ (\ref{eq:beta0}) which can be estimated as $\delta\beta \sim n^{-2}$. Thus a conservative estimate for the validity of our calculations requires $|\gamma| \ll n^{-2}$,
otherwise the degenerate perturbation theory must be used because the difference of the nearest eigenvalues can become smaller than the corrections to the eigenvalues. 

The correlation function to the first order in $\gamma$ and zeroth order in $\epsilon$  may be written as a double sum
\begin{eqnarray}
\label{coefeq1}
\!\!\!\!\!\!\!\!\!\!\!\!\!\!\!\!\!\!\mathbf{C}=\sum_{j,j'}\sum_{\omega,\omega'}\Lambda_{(\omega,j),(\omega',j')}(\underline{\Psi}_{\omega,j}^0+\gamma\underline{\Psi}_{\omega,j}^1)\otimes(\underline{\Psi}_{\omega',j'}^0+\gamma\underline{\Psi}_{\omega',j'}^1)
+ {\cal O}(\epsilon)+{\cal O}(\gamma^2),
\end{eqnarray}
with expansion coefficients calculated from equation (\ref{coefeq})
\begin{eqnarray}
\Lambda_{(\omega,j),(\omega',j')}=\epsilon\frac{(\underline{\Psi}_{\omega,j}^{0*}+\gamma\underline{\Psi}_{\omega,j}^{1*})\cdot \mathbf{Y}' (\underline{\Psi}_{\omega',j'}^{0*}+\gamma\underline{\Psi}_{\omega',j'}^{1*})}{\beta^0_{\omega,j}+\beta^0_{\omega',j'}+\epsilon \beta^1_{\omega,j}+\epsilon \beta^1_{\omega',j'}},
\label{coeflambdaeq}
\end{eqnarray}
where ${\bf Y}=\epsilon{\bf Y}'$ and
\begin{equation}
{\bf Y}'=
\left(
\begin{array}{cccc}
 2 \kappa^{\rm L}_-\sigma^{\rm y}& 0 & \ldots & 0 \\
 0 & 0 & \ldots & 0 \\
 \vdots & \vdots & \ddots &  \vdots \\
 0  & 0 & \ldots & 2\kappa^{\rm R}_-\sigma^{\rm y} 
\end{array}
\right).
\end{equation}
Here an explanation why we used only the environment (hermitian) corrections to the eigenvalues is in order. Let us first consider the unperturbed matrix ${\bf X}_0$ and the anisotropy ($\gamma$) correction ${\bf X}_\gamma$. Both are real anti-symmetric matrices, hence, the eigenvalues of the matrix ${\bf X}_0+\gamma{\bf X}_\gamma$ come in pairs, where one eigenvalue is positive imaginary and the other one is equally imaginary negative. Therefore, by considering only $\gamma$-correction we find singularities in the expression (\ref{coefeq1}), and the condition for the uniqueness of NESS (\ref{conditioneq}) is not satisfied as well. If we include the symmetric (Hermitian), environment part $\epsilon{\bf X}_\epsilon$, anti-symmetry of ${\bf X}$ is broken and the relation (\ref{coefeq1}) becomes regular for all $j,j'$ and $\omega, \omega'$. We conclude that  the environment ($\epsilon$) corrections to eigenvalues are necessary to get a unique NESS. 

Now we will show that indeed only environment corrections to the unperturbed eigenvalues are needed in order to get the correct correlation function when the inequalities $\epsilon\ll\gamma\ll1$ are satisfied. In this regime the coefficients (\ref{coeflambdaeq}) are negligible unless  $\beta_{\omega,j}^0=-\beta^0_{\omega',j'}$,  which is true for $\omega\neq\omega'$ and $j=j'$. In this case also all $\gamma$ corrections to $\beta_{\omega,j}^0$ come in pairs, as discussed before, and thus add up to zero in the denominator of (\ref{coeflambdaeq}), but we have $\beta^1_{1,j}=\beta^1_{2,j}$, which remains the only term in the denominator. Therefore the small parameter $\epsilon$ cancels out of the equation. Hence, only first order environment ($\epsilon$) correction to the eigenvalues is needed to calculate the ``diagonal" coefficients $\Lambda_{(1,j),(2,j)}$, which are the only important contributions to the correlation function, as the rest can be neglected. After simplifying, the largest coefficients are found to be constant (i.e. independent of $j$)
\begin{equation}
\Lambda_{(1,j),(2,j)}=-\Lambda_{(2,j),(1,j)}=: \Lambda_0 = \frac{\kappa_-^{\rm L}+\kappa_-^{\rm R}}{\kappa_+^{\rm L}+\kappa_+^{\rm R}} + \mathcal{O}(\epsilon)+\mathcal{O}(\gamma^2),
\end{equation}
and all the other coefficients vanish in the leading order, i.e. are ${\cal O}(\epsilon^1)$.
The correlation matrix can now be expressed with a single sum
\begin{eqnarray}
\label{corraprox1:eq}
\mathbf{C}\approx\Lambda_0\sum_{j=1}^{n}&\Big(\underline{\Psi}_{2,j}^0\otimes\underline{\Psi}_{1,j}^0 -\underline{\Psi}_{1,j}^0 \otimes \underline{\Psi}_{2,j}^0\\ \nonumber
&+\gamma(\underline{\Psi}_{1,j}^1\otimes\underline{\Psi}_{1,j}^0 +\underline{\Psi}_{2,j}^0 \otimes \underline{\Psi}_{1,j}^1 -\underline{\Psi}_{1,j}^0\otimes\underline{\Psi}_{2,j}^1 -\underline{\Psi}_{1,j}^1\otimes\underline{\Psi}_{2,j}^0)\Big).
\end{eqnarray}
The error of the previous formula is $\mathcal{O}(\epsilon)+\mathcal{O}(\gamma^2)$, thus, factors which involve a direct product of first order corrections are of higher order and are left out. The dyadic products involving zeroth order can be further simplified
\begin{eqnarray}
\label{c2:eq}
\underline{\Psi}_{2,j}^0\otimes\underline{\Psi}_{1,j}^0 -\underline{\Psi}_{1,j}^0 \otimes \underline{\Psi}_{2,j}^0=\sigma^{\rm y}\otimes\left[
\underline{\psi}_j \otimes \underline{\psi}_{j'}\right]
\end{eqnarray}
as well as the product of the zeroth and first order corrections to the eigenvectors\footnote{Note a distinction between {\em dyadic product} of vectors
and Kronecker (tensor) product of matrices, both designated with a symbold $\otimes$, which should be clear from the type of factors.}
\begin{eqnarray}
\label{c1:eq}
\underline{\Psi}_{2,j}^1\otimes\underline{\Psi}_{1,j}^0 +\underline{\Psi}_{2,j}^0 \otimes \underline{\Psi}_{1,j}^1 -\underline{\Psi}_{1,j}^0\otimes\underline{\Psi}_{2,j}^1 -\underline{\Psi}_{1,j}^1\otimes\underline{\Psi}_{2,j}^0\\ \nonumber =\mathrm{i}\sigma^{\rm x}\otimes\sum_{j'=1}^n\frac{K_{j,j'}}{\beta_{1,j}^0+\beta_{1,j'}^0}\left[
\underline{\psi}_{j'}\otimes \underline{\psi}_j - \underline{\psi}_{j}\otimes \underline{\psi}_{j'}\right].
\end{eqnarray}
Summing up the identities (\ref{c2:eq}), (\ref{c1:eq}) and the formula (\ref{corraprox1:eq}) we get the correlation matrix to first order in $\gamma$ approximation
\begin{eqnarray}
\mathbf{C}= \Lambda_0 \sigma^{\rm y}\otimes\mathds{I}_n +\gamma\sigma^{\rm x}\otimes\mathbf{C}^{\rm x}+\mathcal{O}(\epsilon)+\mathcal{O}(\gamma^2),
\label{eq:analyt}
\end{eqnarray}
where
\begin{eqnarray}
\mathbf{C}^{\rm x} := 2\mathrm{i}\Lambda_0\sum_{j,j'=1}^n\frac{K_{j,j'}}{\beta_{1,j}^0+\beta_{1,j'}^0} 
\underline{\psi}_{j'}\otimes\underline{\psi}_j . \label{eq:Cx} 
\end{eqnarray}
The final explicit form of the correlation matrix (\ref{eq:analyt}) with (\ref{eq:beta0},\ref{eq:K},\ref{eq:Cx}) is rather complicated, but it can be simplified for different regimes.

\begin{figure}[htb]
\begin{center}
	\includegraphics[scale=1]{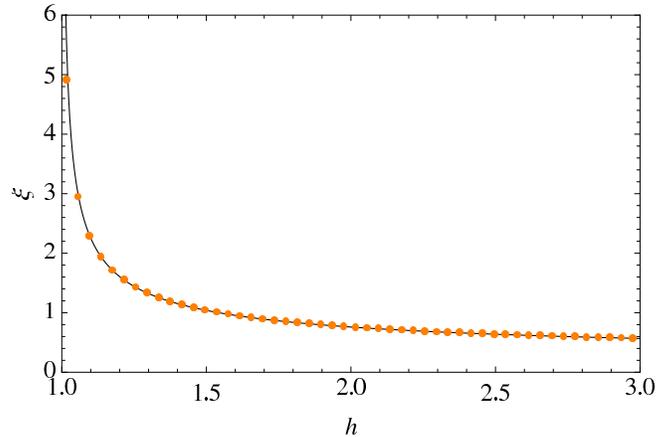}
\caption{Comparison of numerically calculated correlation length $\xi$ (points) with the analytical prediction (\ref{mu_h:eq}) (solid line). In numerical calculations we used a system of size $n=20$, and with parameters $\gamma=10^{-4}$, $\epsilon=10^{-6}.$ Although the size of the chain is not yet very large, agreement of numerical coefficients with analytical prediction is excellent, except perhaps at the left most point where the finite size effect becomes noticable.
In this and all other figures we take bath parameters $\kappa_1^{\rm L}=3, \kappa_2^{\rm L}=1, \kappa_1^{\rm R}=3, \kappa_2^{\rm R}=2$. 
}
\label{mu_h:fig}
\end{center}
\end{figure}

First, we observe that analytic properties of the expression $K_{j,j'}/(\beta^0_{1,j}+\beta^0_{1,j'})$ in (\ref{eq:Cx}) change at $h = h_{\rm c} := 1$. Namely, for $|h| > h_{\rm c}$, the expression (\ref{eq:Cx}) for $\mathbf{C}^{\rm x}$ is analytic for any $n$
and can be evaluated asymptotically for large $n$ giving exponential decay of the correlator
\begin{equation}
C^{\rm x}_{j,j'} \sim {\rm sign}(j-j')\exp(-|j-j'|/\xi), \quad n\to \infty,
\end{equation}
where ${\rm sign}(x>0)=1,{\rm sign}(x<0)=-1,{\rm sign}(0)=0$, with the correlation length
\begin{equation}
\xi= 1/{\rm arcosh}(h) = -1/\ln(h-\sqrt{h^2-1}).
\label{mu_h:eq}
\end{equation}
In fig.~\ref{mu_h:fig}, where we show analytically and numerically calculated correlation lengths $\xi$, one can clearly see vary good agreement of the explicit formula (\ref{mu_h:eq}) with the numerical calculations, and the square-root singularity of the correlation length at the critical field $h=h_{\rm c}$, $\xi \sim |h-1|^{-1/2}$.
 
In the other regime, $|h|<h_{\rm c}$, the formula (\ref{eq:analyt}) can be further simplified in the  poles of the correlation function (or poles of the coefficients $K_{j,j'}/(\beta^0_{1,j}+\beta^0_{1,j'})$), where the above non-degenerate perturbation theory fails. This shall be studied in the following subsection.

The main assumptions that have been made through the calculation are: (i) coupling to the environment is the smallest parameter, so the perturbation correction to the eigenvectors is negligible, while we must consider the corrections to the eigenvalues, (ii) the anisotropy $\gamma$ must be smaller than $\delta\beta$  so non-degenerate perturbation theory
around the isotropic case can be applied.
 
\subsection{Resonances of the correlation function}
In the poles of correlation function the degeneracy of each of the oppositely signed unperturbed eigenvalues $\beta^0_{\omega,j}$, $\omega=1,2$ is twofold. Therefore, we should use the degenerate perturbation theory to determine the correlation matrix. As we shall see, the sole eigenvectors corresponding to the degenerate eigenvalues determine the dominant behavior of the correlation matrix, thus the leading order expression is very simple. 
The unperturbed eigenvalues $\beta_{\omega,k}^0, \beta^0_{\omega,l}$ are degenerate (i.e. $\beta^0_{1,k}=\beta^0_{2,l},\beta^0_{2,k}=\beta^0_{1,l}$) if 
\begin{eqnarray}
h=h_{k,l}:=\frac{1}{2}\left[\cos \left(\frac{\pi k}{n+1}\right)+\cos \left(\frac{\pi  l}{n+1}\right)\right]
\label{eq:res}
\end{eqnarray}
for some $k,l$. 

Let us first assume $k-l$ is {\em odd}, so $K_{l,k}\neq 0$. We then diagonalize $\gamma K_{l,k}\sigma^{\rm x}$ in the degenerate $2\times2$ subspace to find the proper eigenvectors and eigenvalues: 
\begin{eqnarray}
&\underline{\tilde{\Psi}}^0_{1,k}=\frac{1}{\sqrt{2}}(\underline{\Psi}_{1,k}^0-\underline{\Psi}_{2,l}^0),& 
\quad \tilde{\beta}^0_{1,k}=\beta^0_{1,k}-\gamma K_{l,k},\\ \nonumber
&\underline{\tilde{\Psi}}^0_{2,l}=\frac{1}{\sqrt{2}}(\underline{\Psi}_{1,k}^0+\underline{\Psi}_{2,l}^0),& 
\quad \tilde{\beta}^0_{2,l}=\beta^0_{1,k}+\gamma K_{l,k},
\end{eqnarray}
Note again that $\underline{\Psi}_{1,k}^0$ and $\underline{\Psi}_{2,l}^0$ have orthogonal polarizations (\ref{eq:beta0}). Let us assume $\beta_{1,k}>0$ ($k>l$). There is also another degenerate pair with $\beta_{1,l}=-\beta_{1,k}$, $K_{l,k}=-K_{k,l}$ and
\begin{eqnarray}
&\underline{\tilde{\Psi}}^0_{2,k}=\frac{1}{\sqrt{2}}(\underline{\Psi}_{1,l}^0-\underline{\Psi}_{2,k}^0),&
\quad\tilde{\beta}^0_{2,k}=\beta^0_{1,l}-\gamma K_{k,l} = -\beta^0_{1,k}+\gamma K_{l,k},\\ \nonumber
&\underline{\tilde{\Psi}}^0_{1,l}=\frac{1}{\sqrt{2}}(\underline{\Psi}_{1,l}^0+\underline{\Psi}_{2,k}^0),&
\quad\tilde{\beta}^0_{1,l}=\beta^0_{1,l}+\gamma K_{k,l} = -\beta^0_{1,k}-\gamma K_{l,k}.
\end{eqnarray}
All the other eigenfunctions and eigenvalues remain the same $\tilde{\underline{\Psi}}_{\omega,j}^0=\underline{\Psi}_{\omega,j}^0$ and $\tilde{\beta}_{\omega,j}^0=\beta_{\omega,j}^0$, for $j \not\in \{k,l\}$.  As in the non-degenerate case, we only need to calculate the real (dissipative) corrections to the eigenvalues
\begin{eqnarray}
\tilde{\beta}_{\omega,j}^1&=\beta_{\omega,j}^1;\quad\mbox{for }j\not\in \{k,l\},~~ \omega=1,2,\\ \nonumber
\tilde{\beta}_{1,k}^1&=\tilde{\beta}^1_{1,l}=\tilde{\beta}^1_{2,k}=\tilde{\beta}^1_{2,l}=\frac{1}{2}(\beta_{1,k}^1+\beta_{2,l}^1).
\end{eqnarray}
The coefficients $\Lambda_{(1,j),(2,j)}$ on the diagonal are calculated from equation (\ref{coefeq}) with new eigenvectors and eigenvalues and are almost identical to the non-degenerate case
\begin{eqnarray}
\Lambda_{(1,j),(2,j)}=\Lambda_0=\frac{\kappa^-_{\rm L}+\kappa^-_{\rm R}}{\kappa^+_{\rm L}+\kappa^+_{\rm R}},\quad
j\not\in\{k,l\}
\end{eqnarray}
and
\begin{eqnarray}
\Lambda_{(1,k),(2,k)}=\Lambda_{(1,l),(2,l)}&=\frac{\sin ^2\left(\frac{\pi  k}{n+1}\right)-\sin ^2\left(\frac{\pi 
   l}{n+1}\right)}{\sin ^2\left(\frac{\pi  k}{n+1}\right)+\sin
   ^2\left(\frac{\pi  l}{n+1}\right)}\Lambda_0 = \Lambda_0-\delta\Lambda
\end{eqnarray}
where
\begin{equation}
\delta\Lambda := \frac{2\sin ^2\left(\frac{\pi 
   l}{n+1}\right)}{\sin ^2\left(\frac{\pi  k}{n+1}\right)+\sin
   ^2\left(\frac{\pi  l}{n+1}\right)} \Lambda_0 .
\end{equation}
Diagonal coefficients $\Lambda_{(1,k),(2,k)}$ are not constant, therefore, the main contribution to the correlation comes from the leading order in the degenerate subspace
\begin{eqnarray}
&&\underline{\tilde{\Psi}}_{1,k}^0\otimes\underline{\tilde{\Psi}}_{2,k}^0-\underline{\tilde{\Psi}}_{2,k}^0\otimes\underline{\tilde{\Psi}}_{1,k}^0
 + \underline{\tilde{\Psi}}_{1,l}^0\otimes\underline{\tilde{\Psi}}_{2,l}^0-\underline{\tilde{\Psi}}_{2,l}^0\otimes\underline{\tilde{\Psi}}_{1,l}^0 \\ \nonumber
 &&=\sigma^{\rm y}\otimes(\underline{\psi}_l \otimes \underline{\psi}_l - \underline{\psi}_k \otimes \underline{\psi}_k).
 \end{eqnarray}
The correlation matrix (\ref{coefeq1}) has now a simple explicit form
\begin{eqnarray}
\label{cores:eq}
\mathbf{C}_{\rm res} := \mathbf{C}|_{h=h_{k,l}} =\sigma^{\rm y}\otimes(\Lambda_0\mathds{I}_{n}-\mathbf{C}^{\rm y}_{\rm res}) + \mathcal{O}(\gamma) + \mathcal{O}(\epsilon),
 \end{eqnarray}
 where
 \begin{eqnarray}
 \mathbf{C}^{\rm y}_{\rm res}= 2\Lambda_0 \frac{\sin ^2\left(\frac{\pi 
   l}{n+1}\right)  \underline{\psi}_k \otimes \underline{\psi}_k+\sin ^2\left(\frac{\pi 
   k}{n+1}\right) \underline{\psi}_l \otimes \underline{\psi}_l}{\sin ^2\left(\frac{\pi  k}{n+1}\right)+\sin
   ^2\left(\frac{\pi  l}{n+1}\right)}.
   \label{eq:Cy}
\end{eqnarray}
\begin{figure}[!!h]
\begin{center}
	\includegraphics[scale=1]{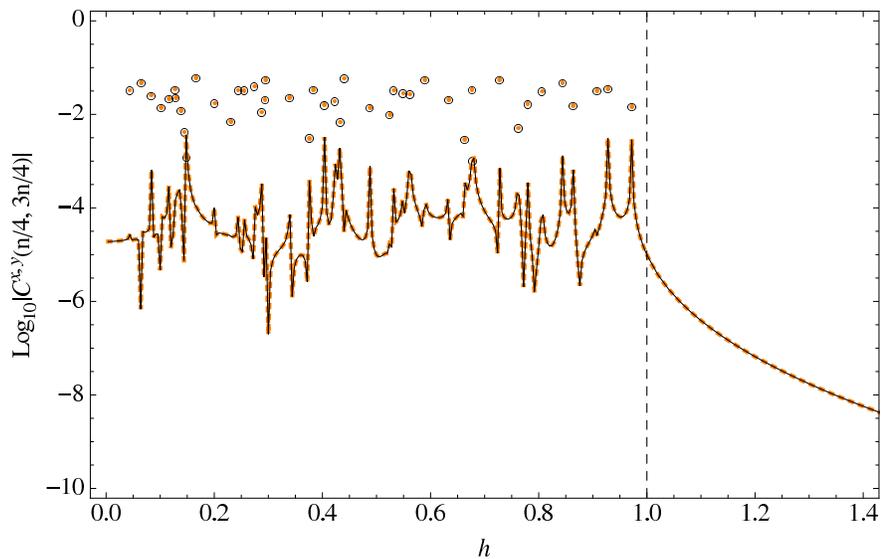}
\caption{Typical correlation matrix element, $C^{\rm x,y}_{n/4,3n/4}$, versus the field strength $h$. In the resonances, $h=h_{k,l}$, the analytical (orange points) and numerical (black circles) correlation element components $C^{\rm y}_{n/4,3n/4}$ are plotted, as the other component $C^{\rm x}_{n/4,3n/4}$ is orders of magnitude smaller. For other values of the field we compare  analytical (orange dashed line) and numerical (black solid line) correlation element components $C^{\rm x}_{n/4,3n/4}$, whereas here the component $C^{\rm y}_{n/4,3n/4}$ is considerably smaller (see fig.~\ref{resonanca:fig}).
In both cases the agreement is excellent. System parameters  are: $n=20$, $\gamma=10^{-4}$, $\epsilon=10^{-6}$.}
\label{spekter:fig}
\end{center}
\end{figure}
Note that this correction is independent of $\gamma$, hence, the correlation in resonance $h=h_{k,l}$ (\ref{eq:res}) scales as $1/n$ - this is hidden in the normalization of $\underline{\psi}_j$ (\ref{eq:beta0}). 
It is also interesting to note that in the leading-order at the resonance the correlation matrix (\ref{eq:Cy}) is a simple combination of two 'sine-waves' and is thus, for large $n$, very much reminiscent of an eigen-mode of a square drum. This is why we choose to call this phenomenon a {\em correlation resonance}.

In order to obtain the next-to-leading first order ${\cal O}(\gamma)$ correction in the resonances we successively apply the degenerate and non-degenerate perturbation theory.
The corrected eigenvectors are then manipulated as in the previous cases resulting in a rather complicated formula
\begin{equation}
\mathbf{C}_{\rm res} =\sigma^{\rm y}\otimes(\Lambda_0\mathds{I}_{n}-\mathbf{C}^{\rm y}_{\rm res}) + \gamma\sigma^{\rm x}\otimes\mathbf{C}^{\rm x}_{\rm res} + 
\mathcal{O}(\gamma^2) + \mathcal{O}(\epsilon),
\label{cores1:eq}
 \end{equation}
 where
\begin{eqnarray}
\label{eq:Cxr1111}
{\bf C}^{\rm x}_{\rm res}&=&2\mathrm{i}\Lambda_0\sum_{1\le j,j' \le n}^{j,j'\neq k,l} \frac{K_{j,j'}}{\beta_{1,j}^0+\beta_{1,j'}^0} 
\underline{\psi}_{j'}\otimes\underline{\psi}_j \\ \nonumber
&&+\mathrm{i}(\Lambda_0-\delta\Lambda)\sum_{1\le j \le n}^{j\neq l}
 \frac{K_{k,j}}{\beta_{1,k}^0+\beta_{1,j}^0} 
(\underline{\psi}_{j}\otimes\underline{\psi}_k -\underline{\psi}_{k}\otimes\underline{\psi}_j )\\ \nonumber
&&-\mathrm{i}(\Lambda_0-\delta\Lambda)\sum_{1\le j \le n}^{j\neq k} \frac{K_{l,j}}{\beta_{1,l}^0+\beta_{1,j}^0} 
(\underline{\psi}_{j}\otimes\underline{\psi}_l -\underline{\psi}_{l}\otimes\underline{\psi}_j )\\ \nonumber
&&+G(\underline{\psi}_{k}\otimes\underline{\psi}_l -\underline{\psi}_{l}\otimes\underline{\psi}_k),
\end{eqnarray}
where the coefficient $G$ is obtained from the restriction of $\mathbf{X}$ in the degenerate subspace.

In the case when $K_{l,k}=0$, which is true for even-even or odd-odd $k,l$, the degenerate eigenvectors are not directly coupled, so the general result is only slightly changed with respect
to the non-degenerate case (\ref{eq:analyt},\ref{eq:Cx}), namely one only needs to exclude the degenerate pair from the double summation
\begin{eqnarray}
\mathbf{C}_{\rm res}= \Lambda_0 \sigma^{\rm y}\otimes\mathds{I}_n +\gamma\sigma^{\rm x}\otimes\mathbf{C}^{\rm x}_{\rm res}+\mathcal{O}(\epsilon)+\mathcal{O}(\gamma^2),
\label{eq:analytr}
\end{eqnarray}
where
\begin{eqnarray}
\mathbf{C}^{\rm x}_{\rm res} = 2\mathrm{i}\Lambda_0\sum_{1\le j,j'\le n}^{(j,j')\neq (k,l),(l,k)} \frac{K_{j,j'}}{\beta_{1,j}^0+\beta_{1,j'}^0} 
\underline{\psi}_{j'}\otimes\underline{\psi}_j . \label{eq:Cxr} 
\end{eqnarray}

\begin{figure}[!!h]
\begin{center}
	\includegraphics[scale=1]{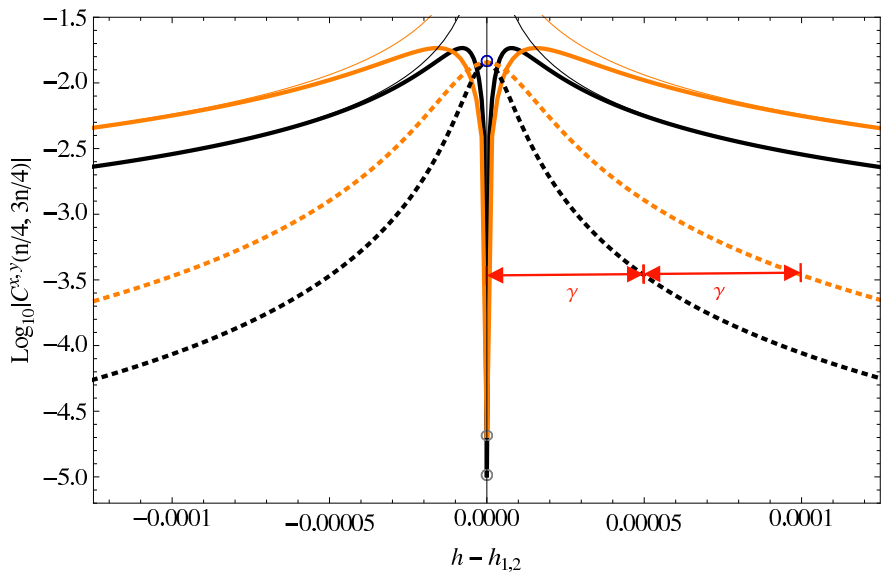}
\caption{Detailed plot of the resonance around $h_{1,2}=(\cos^2(\frac{\pi}{n+1})+\cos^2(\frac{2\pi}{n+1})/2$ for a system with $n=20$, $\epsilon=10^{-6}$ and two values of the anisotropy, $\gamma=0.5\times10^{-4}$ (black curves) and twice larger $\gamma=10^{-4}$ (orange curves). The dashed curves show numerically obtained $C^{\rm y}_{n/4,3n/4}$. The solid curves show $C^{\rm x}_{n/4,3n/4}$, where the thick curves correspond to numerical results and the thin ones to analytical calculation, and  the blue, and gray, circles corresponds to analytically predicted $C^{\rm y}_{n/4,3n/4}$, and $C^{\rm x}_{n/4,3n/4}$, at the resonance field, respectively. The resonance width grows linear with the anisotropy $\gamma$, also the agreement between the (non-resonant) analytically and numerically obtained values of $\sigma^{\rm x}$ projection ($\mathbf{C}^{\rm x}$) is good for $|h-h_{1,2}| > \gamma$.
}
\label{resonanca:fig}
\end{center}
\end{figure}

In fig.~\ref{spekter:fig} we compare the analytical predictions (\ref{eq:analyt}) off the resonances and (\ref{cores:eq}) on the resonances with numerical simulations and find excellent agreement in the entire long range order region $|h| \le 1$. In fig.~\ref{resonanca:fig} we zoom in a narrow window around a single resonance and see very clearly that the resonance width scales as
\begin{equation}
\delta h \approx \gamma,
\end{equation}
i.e. the formula (\ref{eq:analyt}) is essentially applicable for $|h-h_{k,l}| > \delta h$, whereas the formulas (\ref{cores:eq}, \ref{cores1:eq}, \ref{eq:analytr}) hold for $|h-h_{k,l}| < \delta h$.

\begin{figure}[!!h]
\begin{center}
	\includegraphics[scale=1]{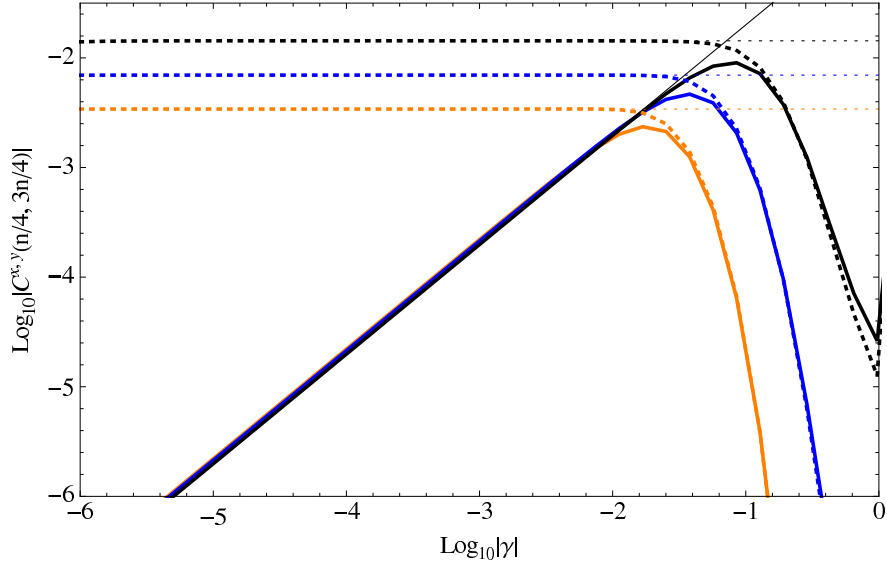}
\caption{Comparison between analytical expression (thin lines) and numerical calculations (thick curves) for the correlators $C^{\rm x}_{n/4,3n/4}$ (solid lines) and $C^{\rm y}_{n/4,3n/4}$ (dotted lines) versus the anisotropy parameter $\gamma$. Results for $n=20$ (black),  $n=40$ (blue), $n=80$ (orange) are plotted. Note that results show clearly that perturbative analytics accurately describes the numerics up to $\gamma \approx 1/n$. Horizontal thin lines are calculated from the equation (\ref{eq:Cy}) and the diagonal (dashed) thin line from the formula (\ref{eq:Cxr1111}).
} 
\label{gamma_dep:fig}
\end{center}
\end{figure}

Before closing, several interesting remarks are in order. Firstly, we note that in resonances the correlation function $\mathbf{C}^{\rm y}$ (\ref{eq:Cy}) does not depend on the anisotropy $\gamma$, however
the use of perturbative arguments indicates that this is true only for small $\gamma$, i.e. $|\gamma| < \delta\beta$.
As noted before, for larger but still small anisotropies $(1/n^2\ll\gamma\ll1)$ degenerate perturbation theory must be used as some pair of unperturbed eigenvalues
$\beta^0_{\omega,j}$ may become nearly degenerate.  In other words, for fixed small anisotropy, our results are quantitatively valid only up to a maximal size $n_{\rm max} \sim |\gamma|^{-1/2}$.  However, numerical calculations (see fig.~\ref{gamma_dep:fig}) indicate that this estimate is too conservative, in fact
$\gamma$ can be as large as $\sim 1/n$ for our results to remain valued. This means that, in practice, $\delta\beta$ can be estimated as an {\em average spacing} between eigenvalues (\ref{eq:beta0}), rather than the more conservative {\em minimal spacing}.
Nevertheless, the fact that we cannot extend our calculation to thermodynamic limit $n\to\infty$ for any fixed system parameters implies that our method can not be used to describe the critical closing of Liouvilean spectral gap behavior as observed in \cite{njp,njp2}.

\begin{figure}[!!h]
\begin{center}
	\includegraphics[scale=1]{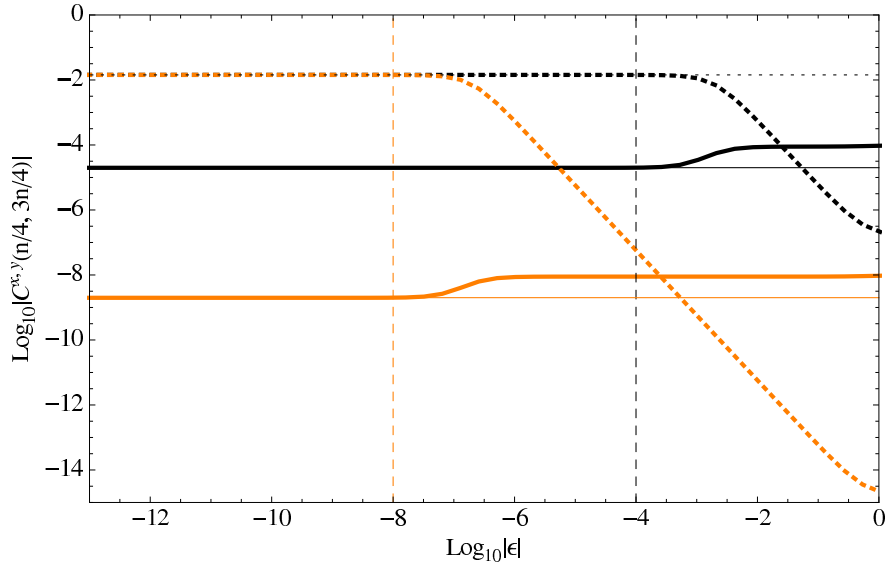}
\caption{Comparison between analytical expression (thin horizontal lines) and numerical calculations (thick curves) for the correlator $C^{\rm x,y}_{n/4,3n/4}$ 
versus the reservoir coupling strength $\epsilon$, and for two different values of anisotropy $\gamma=10^{-4},10^{-8}$ (black, orange, respectively) indicated with vertical dashed lines. Solid lines refer to ${\bf C}^{\rm x}$ component and dashed lines to ${\bf C}^{\rm y}$ component of the correlation matrx.} 
\label{eps:fig}
\end{center}
\end{figure}

Secondly, we stress again an interesting fact that for small reservoir coupling strength $\epsilon$ our leading order perturbative results do not depend on $\epsilon$, meaning for example that the value of the spin-spin correlation at long distance in the regime $|h| < h_{\rm c}$ is insensitive to $\epsilon$ even for infinitesimal coupling, but there is, however, dependence on the properties of the baths (e.g. effective temperature, chemical potential, etc) as encoded in parameter $\Lambda_0$.
We note that in the models of symmetric out-of-equilibrium driving at infinite temperature (e.g. \cite{znidar,prosaito,prozni10}), where formally $\Lambda_0=0$, one obtains a different result: $\mathbf{C} = {\cal O}(\epsilon)$. With increasing $\epsilon$, our results remain valid - due to our perturbative assumptions - up to $\epsilon^* \sim \gamma$. This is illustrated with simple numerical calculations in fig.~\ref{eps:fig}.

The last point we make concerns the total number $N(n)$ of resonant values of the magnetic field $h_{k,l}$ for a given chain length $n$. A simple inspection of the formula (\ref{eq:res})
with the constraint that $k-l$ should be {\em odd}, results in an expression for the resonance counting number
\begin{equation}
N_{\rm res}(2n)=N_{\rm res}(2n-1)=n (n-1).
\end{equation}
If we consider only the values of the field with a fixed, say positive sign, $h>0$, then the number of resonances is $N(n)/2$.
This amounts to, for example, in total of $45$ resonances shown in fig.~\ref{spekter:fig} for $n=20$.

\section{Discussion and conclusion}
The present article contains two main results. Firstly, using the quantization in the Fock space of operators (`third quantization') we have derived manifestly bilinear form of the many-body Liouvillean and expressed the equation of motion for the 2-point correlation function in terms of the continuous Lyapunov equation.
Secondly, we have presented explicit analytical result for the steady state correlations in boundary driven open XY spin 1/2 chain in the Lindblad formulation.
The two regimes of short range and long range correlations have been clearly identified, with the non-equilibrium phase transition occurring at the critical value of transverse magnetic field $h_{\rm c} = 1$. In the short range regime, exponent of correlation decay has been obtained analytically, whereas in the regime of long range order, a particular resonant spikes of the correlation response of the system have been identified, where the values of the correlator (scaling with the inverse of the chain length $n$) have been calculated in a simple closed form. 

We note that our analysis could perhaps be extended to the case of large (or better to say, non-small) anisotropies, where the continuous limit (for $n\to\infty$) of the Lyapunov equation of the XY model becomes a non-homogeneous Helmholtz equation driven by two point-like sources representing the two Lindblad reservoirs. 
The interpretation of the non-equilibrium phase transition in this picture becomes rather obvious, namely long-range order corresponds to positive energies and real wave numbers, whereas short range phase correspond to negative energies and imaginary (evanescent) wavenumbers. 
The correlation resonances can then be exactly identified with the eigenstates of a square-shaped quantum billiard (with an appropriate symmetry/boundary contition).
Nevertheless, the exact analytical results for the correspondence between the open (strongly anisotropic) XY chain and the driven Helmholtz equation remains a work in progress.

We hope that our results, which are nontrivial though surprisingly simple, may be relevant and observable in real laboratory experiments.
 
\section*{Acknowledgements}
We acknowledge financial support by the Programme P1-0044, and the Grant J1-2208, of the Slovenian Research Agency (ARRS).

\end{document}